# Twisted Helical shaped Graphene Nano-Ribbons: Role of Symmetries and Passivation

Rajesh Thakur*[$], P. K. Ahluwalia[$], Ashok Kumar[&], Munish Sharma and Raman Sharma[$]

[$]*Department of Physics, Himachal Pradesh University, Summer Hill, Shimla 171005 Himachal Pradesh (INDIA)*
[&]*Department of Physical Sciences, School of Basic and Applied Sciences, Central University of Punjab, Bathinda, 151001 Punjab (INDIA)*
[#]*Department of Physics, School of Basic and Applied Sciences, Maharaja Agrasen University, Atal Shiksha Kunj, Barotiwala, 174103, Himachal Pradesh (INDIA)*

*Corresponding author: Email: rajeshhputhakur@gmail.com

**Abstract**

The Hydrogen and Fluorine planar armchairs graphene nanoribbons (H **&** F AGNRs), subjected to twist deformation within fixed periodic boundary conditions, eventually morph to a helical conformations are investigated at few tractable points. Unlike structural properties, no effect of symmetries on mechanical properties is observed, though passivation does have a significant effect on mechanical as well as on electrical properties. Hooke's law for severely twisted AGNRs indicates the high elasticity of H-AGNRs whereas the F-AGNRs shows plasticity after threshold torsional strain. Torsional stress $(E_\theta)$ is approximated from the variation in total energy $(\Delta E)$ with square of torsional strain $(\theta^4 \Sigma^4)$. Further, the effect of passivation on the electronic properties of helical conformations with different torsional strain is decisive in metal-to-semimetal and semimetal-to-metal transition. The band gap response of narrow GNRs **N**=6, 7 & 8, within a fixed cell under sever twisting arranged itself in two group as $(i)$ monotonously increasing for $q = 0,2$ and $(ii)$ decreasing for $q = 1$, here $q = mod(N, 3)$ in effective strain space $(\theta^2 \Sigma^2)$. This trend has also been observed for Fluorine passivated AGNRs, though band gap of **N**=7 F-AGNRs drops from $E_g \simeq 0.95$eV to $E_g \simeq 0.05$eV at extreme torsional strain forming Dirac cone at $\pm K$ allows dissipation less transport to charge carriers of high kinetic energy at low bias.

1. **Introduction**

The strain engineering is one of the most efficient way to tune the electronic properties of nanostructures. The various ways to apply strain in graphene nanoribbons (GNRs) including bending[1,2], twisting[3] and intrinsic rippling[4] can be used to develop materials with tunable interaction of core and edge regions beyond simple straintronics[5,6]. One can also anticipate possible influence on the electronic properties of GNRs by applying an asymmetric strain along the width of ribbons. Furthermore, bending and twisting break the mirror and inversion symmetry[7], hence, numerous desirable properties of GNRs which are likely to be influenced still remain unexplored under such conditions, which may have new possibilities and applications in various technologies and devices[8]. Moreover, the chemistry at the periphery atoms of the GNRs is most influential in modulating the band gap of narrow GNRs which has prompted experimentalist to demonstrate that these GNRs can be used as molecular wires which exhibit metallic behavior even at room temperature[9] leading to their usage as ideal interconnects in molecular scale electronic circuitry.

The developments of GNR-based nano-electronics[10] needs further impetus to bring it out of its seemingly saturated possibilities and this provides us a strong motivation to investigate the topologically helical shaped soft conformations of GNRs from symmetry perspective. Note that helical conformations can either be self-assembled or be fabricated by several growth and fabrication techniques that have been successfully used to produce different chiral systems[11–13]. Recently, a range of synthesis methods has been developed to obtain twisted GNRs[14–17], which opens up a new possible route to potential applications of chiral systems including THz generation[18–21], stretchable electronics[22,23], non-linear electronics[24–26] and spin selectivity[27–31].

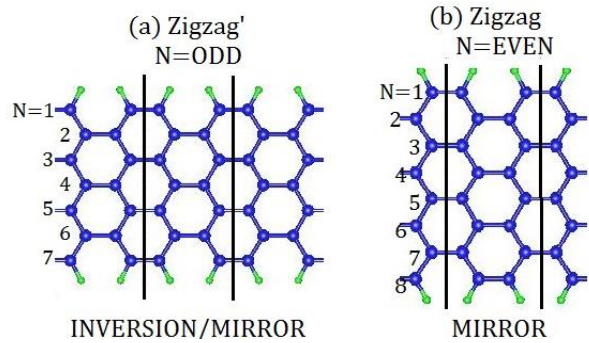

**Figure 1:** *Categorization of structural topology of untwisted AGNRs. The nanoribbons are identified according to the type of (a) zigzag' termination having inversion & mirror symmetry and (b) zigzag termination having only mirror symmetry respectively. Widths that correspond to value of number of dimer (N) along the lateral direction are labeled from "1" to "N". Schematics of AGNR unit cell structures are shown with hydrogen (Fluorine) passivation. The unit cell of each structure that is commensurate with the termination is indicated by the bold black vertical lines.*

The mechanical and electronic properties of Armchair graphene nanoribbons (AGNRs) are sensitive not only to their width and edge chemistry but also to the morphology. Therefore, in this paper, we study the effect of twisting and passivation of Hydrogen and Fluorine on structural, mechanical and electronic properties of a twisted helical of AGNRs[3,32–34] within the fixed boundary condition making translational symmetry tractable for a few discrete twist angles values ($\theta$), which have not been explored to the best of our knowledge. AGNRs with zigzag' type termination with **N**=odd (Figure 1(a)) have both mirror and inversion symmetry, however, AGNRs with **N**=even (Figure 1(b)) have only mirror symmetry. The reason to investigate the GNRs of narrow width is that it favors twists and forms the helical shaped morphologies with small size. It is to be mentioned that the use of periodic boundary condition (PBC) for helical shaped GNRs allows only discrete torsional deformations compatible with the translational symmetry along Z-direction, opens up novel route to inquire and extract the physics of 1D twisted ribbons in the nature.

In this work we have introduced a novel way to define a lattice constant of super cell of helical AGNRs which formed on twisting a one dimensional (1D) untwisted super cell and applicable for all 1D systems. After computational details (Section 2), we will find the answer of following questions in the rest of the article:

(A)In section (3.1): How does symmetry effect the super cell lattice of helical AGNRs? (B) In section (3.2): What would be the response of GNRs' mechanical properties to twist and what if we passivate the AGNRs with different type of atoms? What could be the possible reason which alters its mechanical properties? Is there any mathematical model that can relates the mechanics of AGNRs passivated with different atoms with their elasticity? (C) In section (3.3): What happened to the electronic band gap when we twist the helical AGNRs with different torsional angle? Is there any effect on band gap if we change the edge environment by changing the passivated atom? How can we define the dependence of band gap with torsional angle or torsional strain mathematically? Does the curvature of AGNRs influences the capacity to hold the electron density of AGNRs? (D) In Section (3.4) we have made comparison of our results with the previous studies and finally in Section (4) we give the summary of the work we have done.

## 2. Computational details

All the calculations presented in this work are performed using spin-polarized first principles method by using Spanish Initiative for Electronic Simulations with Thousands of Atoms (SIESTA)[35] simulation package. Normconserving Troullier Martin pseudopotential in fully separable Kleinman and Bylander form has been used to treat the electron-ion interactions[36]. The exchange and correlation energies have been treated within GGA-PBE functional[37]. The Kohn Sham orbitals were expanded as a linear combination of numerical pseudo atomic orbitals using a split-valence double zeta basis set with polarization functions (DZP). The convergence for energy is chosen as $10^{-5} eV$ between two electronic steps.

Throughout geometry optimization, the confinement energy of numerical pseudo-atomic orbitals is taken as 0.01Ry. Minimization of energy was carried out using standard conjugate-gradient (CG) technique. Converged values of sampling for the $k$-mesh grid $\sim 10^{-2} Å^{-1}$ have been used according to Monkhorst-Pack scheme[38] to sample Brillioun zone.

Furthermore, without allowing the axial relaxation of cell the helical structures were relaxed until the force on each atom was less than $10^{-2} eV Å^{-1}$, it provides an effective way to explore both axial and torsional deformation simultaneously. The spacing of the real space used to calculate the Hartree exchange and correlation contribution of the total energy and Hamiltonian was 800Ry for untwisted AGNRs. The converged values were taken in a range between 1300Ry to 1450Ry for twisted AGNR. Vacuum region of about ~12Å along X and Y-direction was used in calculations to prevent the superficial interactions between the periodic images.

To check the reliability and the accuracy of numerical orbitals basis sets, a test calculations for untwisted N=4 GNR along zigzag direction using plane wave basis sets code VASP[39,40] are performed (See Figure S1) for comparison, which are in very good agreements in terms of curvature matching of bands.

### 3. Results and Discussion

The calculated lattice parameters for hydrogen passivated AGNRs (H-AGNRs) unit cells are found to be in good agreement with the previously reported values obtained using density functional theory (Table 1&2). The lattice constants for $N$=6 H-AGNRs, N=7 H-AGNRs and N=8 H-AGNRs are found to be 4.33Å, 4.32Å and 4.31Å, respectively (Figure S2(a)). We multiply the unit cell to form a super cell and twist the AGNRs mechanically by $\pi$-rad to form helical morphologies. Because of inverse Poynting effect[41], a significant stress was found along the AGNRs axis. In this study we restrict ourselves only on those helical shaped systems whose lattice constants are described by symmetry with fixed end results in axial strain, hence, experiencing torsional and tensile strain simultaneously.

Similarly the calculated lattice constant of fluorine passivated AGNRs (F-AGNRs) for **N**=6, 7 and 8 is $L_o$=4.44 Å, 4.42 Å and 4.41Å, respectively (Figure S2(b)).

### 3.1. Structural Properties

We classified different helical conformations by four parameters: lattice constant ($L_M$), torsional angle($\theta$), width (**W**), torsional strain ($\gamma_{max}$) that depends on width of AGNRs as given in the Table 1 for **N**=even and Table 2 for **N**=odd. The subscript **M** of lattice constant $L_M$ is the number of times unit cell get repeated or multiplied to have required twisted super cell. To obtain the helical configuration, the system is subjected to a mechanical force to artificially twist the ribbon from the flat configuration into a helical conformation as described in Figure 2((a) to (c)). In order to incorporate boundary conditions the torsional angle is calculated in a way that the positions of atoms do not lose their periodicity and always coincide with the positions of atoms in the next super cell even after 180° rotation. Thus the lattice constants for twisted unit cell follow the relation:

$$L_M = M * L_0 \qquad (1)$$

Here **M** is the multiplicity factor defined as $M = I + \frac{S}{2}$ as **I** is any integer and for **N**=odd AGNRs the **S**=0 which do not possess inversion symmetry and for **N**=even AGNRs the **S**=1 which do possess inversion symmetry as depicted in Figure 1. Therefore, we may classify the helical conformations in two groups $S = mod(N, 2)$ is defined by the inversion symmetry results into half-integral and integral multiplication of unit cell to form a periodic super cell of 1D helical conformations.

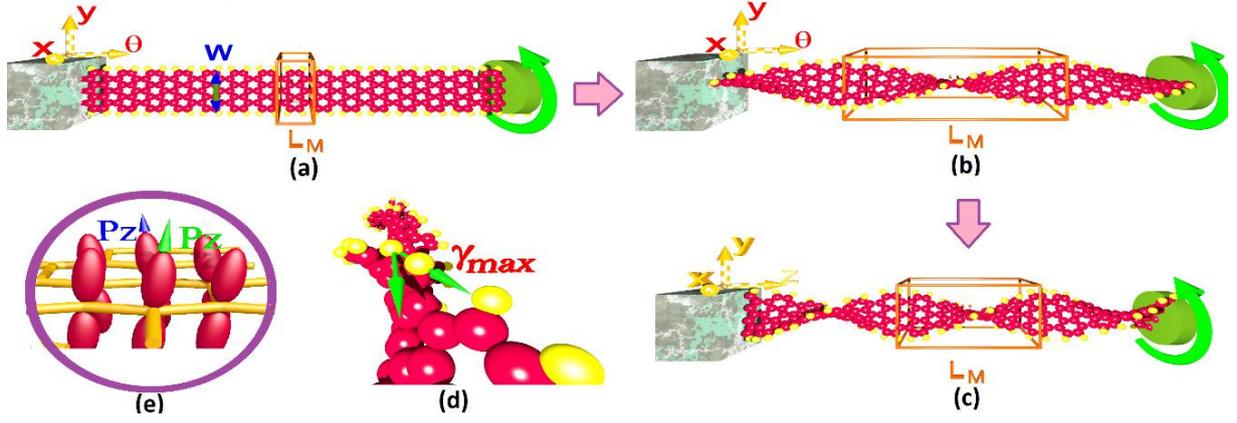

**Figure 2:** *Schematic description of the mechanics of twisting that has been applied to an infinitely long (a) planar N=8 AGNR having minimum primitive lattice constant $L_M$=4.32Å and torsional angle θ=zero that morphs it into (b) helical shaped infinitely long twisted AGNR, now with increased value of lattice constant $L_M$=37.21Å and torsional angle θ=0.121rad that was further twisted to (c) a helical of smaller lattice constant $L_M$ and larger torsional angle θ. Torsional angle increased in ascending order as for fig(a) < fig(b) < fig(c). In this way we have the few tractable arrangements of twisted ribbon that meet with the translational symmetry constraints. (d) $γ_{max}$ is the angle made by the tangent of edge atoms with helical axis θ (green arrow pointing outward to the plane of paper). (e) The misalignment of $P_Z$ Orbitals of two adjoining C-C atoms bonds under twisting.*

We are interested in studying the non-uniform planar tension situation which arises in the twisted AGNR and is different for different values of torsional angle. Cylindrical coordinates are chosen as these are most appropriate to study the torsional effect. If we take Z-axis as a AGNRs axis then the torsional strain ($γ$) along that actually gets induced, as a result of torsional angle ($θ$) coincident with AGNRs axis, can be expressed as:

$$γ = ρ \frac{dθ}{dz} \quad ; 0 ≤ ρ ≤ W/2 \qquad (2)$$

The $ρ$ is a distance of a point from the axis of rotation perpendicular to Z-axis, $W$ is the width of ribbon, $dθ$ is the torsional angle and $ρ_{max} = W/2$ for twisted AGNR. For any twisted AGNR, the $dz$ equals $L_M$ and $dθ$ equals $π$ rad. The value of shear strain is maximum at the peripheral position of ribbon which can be expressed as,

$$γ_{max} = ρ_{max} \frac{dθ}{dz} \qquad (3)$$

This equation gives rise to the monotonous increasing strain towards the edges from the axis of the center of rotation and leads to a maximum strain along the peripheral position of GNRs and no strain along the AGNRs axis.

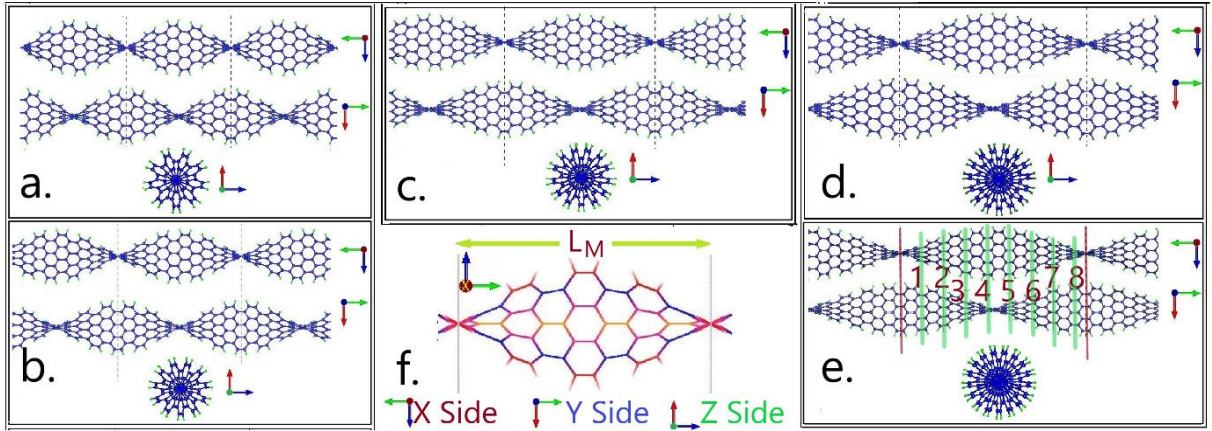

**Figure 3**: *Twisted N=8 H-AGNRs with multiplicity M, torsional angle (θ) and lattice constant for (a) 4.5, 0.09rad & 36.72Å (b) 5.5, 0.10rad & 32.33Å (c) 6.5, 0.11rad & 28.02Å (d) 7.5, 0.13rad & 23.71Å and (e) 8.5, 0.16rad & 19.40Å, respectively. Black dashed lines are the periodic boundary condition for the twisted AGNR super cell. In (e) we elaborated the value of $M = S/2 + I$ for I=8 twisted 8-AGNR, that comes out to be 8.5 and written as a subscript for each lattice constant value in Table 1. (f) Depicts N=7 H-AGNRs for M=4.0, with each triad point is a C atom and lines joining these points are C-C bonds. Colors represent the strain linearly depends on the distance from axis of AGNR (Equation 2). Yellow triad points lies on the axis of twisted ribbon. Gray solid lines are the lattice constant $L_M$.*

### 3.1.1. N=even Twisted AGNRs

Even twisted (**N**=6 and 8) AGNRs have even number of dimer lines; therefore, these two do not have inversion symmetry Figure 1(b). For further elaboration we take **N**=8 H **&** F AGNRs. The lattice constant $L_M$ of H and F AGNRs for minimum multiplicity value **M**=4.5 and **S**=1 is 19.40Å **&**19.85Å , respectively, which corresponds to the most twisted nanoribbons within a fix periodic boundary as shown in Figure 3(a). Hence, these have maximum torsional angle $\theta$=0.16$rad\text{Å}^{-1}$ for both, torsional strain $\gamma_{max}$=0.68$rad$ **&** 0.66$rad$ for H & F AGNRs, respectively, and the narrowest width 8.42Å **&** 8.35Å, respectively, among all the twisted 8-AGNRs we have studied. Figure 3(a-to-e) represent a ball and stick model for twisted **N**=8 H-

**Table 1** *Calculated structural parameters for the edge passivated N=even AGNRs for different values of torsional angle($\theta$). The $L_M$ is a lattice constant of the cell and the subscript M of lattice constant $L_M$ is the number of times unit cell get multiplied to have required twisted super cell, $\gamma_{max}$ is the maximum torsional strain and the width of AGNRs is denoted by W. Theoretical [42]@, [43]# and [44]$*

| System | $L_M$(Å) | $\theta(rad Å^{-1})$ | $\gamma_{max}(rad)$ | $W$(Å) |
|---|---|---|---|---|
| 6HAGNR | $4.33_{1.0}$ | Zero | Zero | 6.15, 6.15#, 6.19$ |
| | $32.44_{7.5}$ | 0.10 | 0.30 | 6.13 |
| | $28.11_{6.5}$ | 0.11 | 0.34 | 6.12 |
| | $23.79_{5.5}$ | 0.13 | 0.40 | 6.11 |
| | $19.46_{4.5}$ | 0.16 | 0.49 | 6.09 |
| | $15.14_{3.5}$ | 0.21 | 0.63 | 6.03 |
| 8HAGNR | $4.31_{1.0}$, $4.35$@ | Zero | Zero | 8.63, 8.67#, 8.66$ |
| | $36.72_{8.5}$ | 0.09 | 0.37 | 8.57 |
| | $32.33_{7.5}$ | 0.10 | 0.42 | 8.56 |
| | $28.02_{6.5}$ | 0.11 | 0.48 | 8.54 |
| | $23.71_{5.5}$ | 0.13 | 0.56 | 8.49 |
| | $19.40_{4.5}$ | 0.16 | 0.68 | 8.42 |
| 6FAGNR | $4.44_{1.0}$ | Zero | Zero | 6.07 |
| | $33.32_{7.5}$ | 0.09 | 0.29 | 6.06 |
| | $28.88_{6.5}$ | 0.11 | 0.33 | 6.05 |
| | $24.44_{5.5}$ | 0.13 | 0.39 | 6.05 |
| | $19.99_{4.5}$ | 0.16 | 0.47 | 6.02 |
| | $15.55_{3.5}$ | 0.20 | 0.60 | 5.97 |
| 8FAGNR | $4.41_{1.0}$ | Zero | Zero | 8.54 |
| | $37.49_{8.5}$ | 0.08 | 0.36 | 8.51 |
| | $33.08_{7.5}$ | 0.10 | 0.40 | 8.49 |
| | $28.67_{6.5}$ | 0.11 | 0.46 | 8.47 |
| | $24.26_{5.5}$ | 0.13 | 0.55 | 8.42 |
| | $19.85_{4.5}$ | 0.16 | 0.66 | 8.35 |

AGNRs having value of torsional strain ($\theta$) in descending order and $M$ is in ascending order from (a) $M$=4.5 < (b) $M$=5.5 < (c) $M$=6.5 < (d) $M$=7.5 and (e) $M$=8.5. In Table 1 we have given the calculated lattice parameters for $N$=6 and 8 (i.e. even) for H & F passivated AGNRs. Because of the absence of inversion symmetry the value of multiplicity factor ($M$) comes out to be half integral as explained pictorially in Figure 3(e). Torsional strain is in ascending order from $M$=8.5 to $M$=4.5, thus Beyond $M$=4.5 the AGNRs broke off and hence were not considered for the present study (See Figure S3).

### 3.1.2. N=odd Twisted AGNRs

In the case of odd twisted AGNRs, due to inversion symmetry (Figure 1(a)), the $S$=0 and the values of $M = 4 \leq M \leq 8$ in Equation (1) gives the value of lattice constant of twisted helical H-AGNRs within periodic super cell. The lattice constant of most twisted $N$=7 H & F AGNRs is $L_M$=17.26Å & 17.68Å and the width ($W$) of the ribbons is 7.13Å & 7.04Å, respectively. The value of torsional angle ($\theta$) and maximum torsional strain $\gamma_{max}$ keeps on increasing as the value goes from $M$=8 to $M$=4 as given in the Table 2. Figure 3(f) shows the wire-straw model for twisted $N$=7 AGNRs, with each triad point representing C atom and the lines connecting these atoms are C-C bonds. Strain depends linearly on the distance of dimer lines from AGNRs axis (Equation 2) is differentiated by color code as it increased in ascending order as Yellow < Magenta < Blue < Red. Because of odd number of dimer lines yellow triad points lies on the axis of twisted ribbon. In Table 2 the lattice parameters for remaining torsional angle for helical shaped AGNRs unit cell for $N$=7 H & F AGNRs are listed.

From above discussion, we can conclude that for bipartite AGNRs the symmetry has played a decisive role in defining the unit cell lattice parameters of helical shaped AGNRs. On application of torsional strain the mirror and inversion symmetry get broken. Note that only within fixed end the AGNRs' width kept decreasing and eventually get broken. Narrower

AGNRs were twisted up to higher torsional angle than wider AGNRs because of strain experienced by the edge atoms was higher in case of wider AGNRs.

Table 2 *Calculated structural parameters for the edge passivated N=7 (i.e. odd) AGNRs for different values of torsional angle($\theta$). The $L_M$ is a lattice constant of the cell and the subscript M of lattice constant $L_M$ is the number of times unit cell get multiplied to have required twisted super cell, $\gamma_{max}$ is the maximum torsional strain and the width of AGNRs is denoted by W. Theoretical [43][#] and [44][$]; Experimental [45][&]*

| System | $L$(Å) | $\theta(rad Å^{-1})$ | $\gamma_{max}(rad)$ | $W$(Å) |
|---|---|---|---|---|
| 7HAGNR | $4.32_{1.0}$, $4.2^{\&}$ | Zero | Zero | $7.37$, $7.31^{\#}$, $7.34^{\$}$, $7.4^{\&}$ |
|  | $34.53_{8.0}$ | 0.09 | 0.33 | 7.32 |
|  | $30.21_{7.0}$ | 0.10 | 0.38 | 7.31 |
|  | $25.89_{6.0}$ | 0.12 | 0.44 | 7.27 |
|  | $21.58_{5.0}$ | 0.15 | 0.53 | 7.22 |
|  | $17.26_{4.0}$ | 0.18 | 0.65 | 7.13 |
| 7FAGNR | $4.42_{1.0}$ | Zero | Zero | 7.26 |
|  | $35.37_{8.0}$ | 0.09 | 0.32 | 7.24 |
|  | $30.95_{7.0}$ | 0.10 | 0.37 | 7.22 |
|  | $26.53_{6.0}$ | 0.12 | 0.43 | 7.19 |
|  | $22.11_{5.0}$ | 0.14 | 0.51 | 7.14 |
|  | $17.68_{4.0}$ | 0.18 | 0.63 | 7.04 |

### 3.2. Mechanical Properties: Twist and Tension on GNRs

It has been reported previously that inverse Poynting effect[41] (shrinking of nanoribbons on twisting along AGNRs axis) triggered by an axial tensile stress $\sigma$ (along Z-direction) is associated with the twist-induced inhomogeneous tensile strain. Since we are considering twisting within the fixed end, the resultant strain in AGNR axis induces Poynting Stress, expressed in Voigt tensor form (Table S1) which has only one non-zero component out of six: [σx, σy, σz, σxy, σxz, σyz] ⇒ [0, 0, σz, 0, 0, 0] and this was calculated for all considered systems. Positive Poynting stress value shows the tensile strain that been experienced by the twisted H & F AGNRs, and it confirms the case of inverse Poynting vector of these systems.

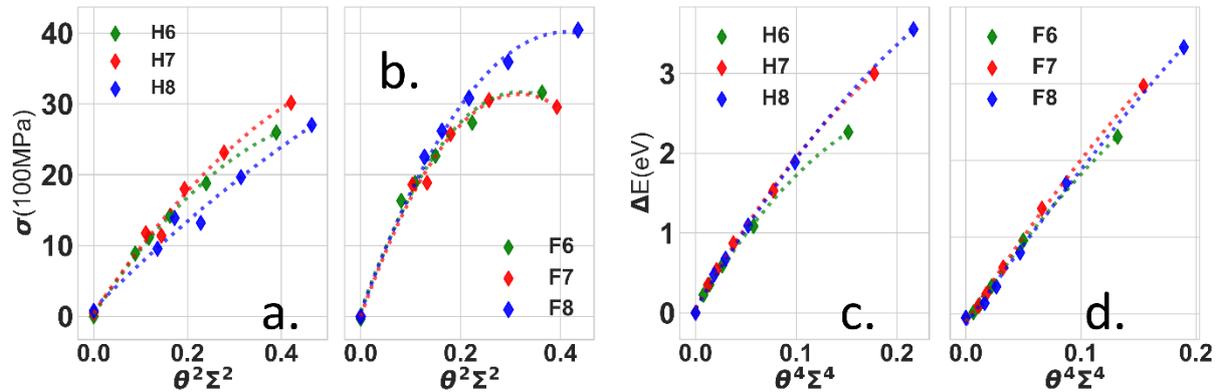

**Figure 4:** *(a) and (b) Computed axial Poynting Stress ($\sigma_z$) versus $\Sigma^2\theta^2$ under fixed boundary condition for H & F AGNRs, respectively. Calculated total energy change per unit primitive cell $\Delta E$ versus $\Sigma^4\theta^4$ validated the approximation even for severely twisted (c) H-AGNRs and (d) F-AGNRs.*

The stress under the induced effective tensile strain $\varepsilon_n$ is obtained with Hooke's law as $\varepsilon_n$ proportional to $\sigma_z$. Furthermore, within the elastic approximation $\sigma_z \propto \varepsilon_n$ where $\varepsilon_n = \frac{1}{2}\theta^2\Sigma^2$ i.e. effective tensile strain associated with the twist angle induces misalignment angle (See Figure 2(e)) in the adjoining C-C bonds. Here we propose the $\Sigma^m = 2\int_0^{W/2} \rho^{m-1} d\rho$, where the width ($W$) of the ribbons at each twist is given in Table 1&2. Range of limit dictates the symmetry that arises from the coincidence of torsional strain and AGNRs axis. The tensile component of energy ($E_s$) follows from relation $E_s \simeq \frac{1}{2}\frac{C}{(1-\nu^2)}\varepsilon_n^2$ as:

$$E_s \simeq \frac{1}{8}\frac{C}{(1-\nu^2)}\theta^4\Sigma^4 \qquad (4)$$

Here $C$ is the stiffness constant and $\nu$ is the Poisson's ratio, and $\theta$ is the torsional angle. The analysis of the stress-strain relation behavior for severe torsional strain has been done by plotting $\sigma_z$ (axial Poynting stress) versus $\theta^2\Sigma^2$ in Figure 4(a&b) for H & F passivized AGNRs, respectively. The linear dependence of $\sigma_z$ on $\theta^2 \Sigma^2$ corresponds to the elastic deformation. In case of F-AGNRs the higher order terms account for both nonlinear elastic deformation and strain softening[46]. The $\sigma_z - \Sigma^2 \theta^2$ curves shows linear variation for H-passivated AGNR indicating unusually high flexibility (i.e. the linear nature of stress and strain dependence) prevails even for higher torsional strain. In contrast, F-AGNRs at lower range of $\theta$ taken between $0.00 < \theta < 0.110$, $\sigma_z$ is a linear function of $\Sigma^2 \theta^2$ while the higher order terms of Taylor's series are neglected. Therefore, in this range the variation can be regarded as Harmonic in nature for all F-AGNRs. On further increase in torsional strain, the contribution of nonlinear terms is indicating the inharmonic variation implies that the deformation is plastic in this region.

The loss of linear behavior for higher torsional strain values in F-AGNRs indicating some threshold for the extent of elongation of C-C bonds. Note that misalignment angle between the two π-orbitals located on two connected C atoms along the AGNR axis depends on $\theta$ (See Figure 2(e)). Figure 4(b), the loss of linear behavior for higher torsional strain values in F-AGNRs indicating some threshold for the extent of elongation of C-C bonds. Note that misalignment angle between the two π-orbitals located on two connected C atoms along the AGNR axis depends on $\theta$ (See Figure 2(e)). Within elastic linear behavior for lower torsional angle $\theta$, the stiffness i.e. slope of lines, is comparatively higher than H-AGNRs and therefore, loss in its elastic nature for severely twisted deformation, though both behave in flexible manner in the same range of induced stress.

In order to get further insights into the deformation of AGNRs, we decompose $\Delta E$ energy as a sum of tensile strains energy and torsional strain energy as $\Delta E = E_s + E_\theta$. Interestingly, in Figure 4(c & d), we found a linear correspondence between the changes in energy per unit cell ($\Delta E$) and $\theta^4 \Sigma^4$ showing almost linear character for all H & F AGNRs. The linear character support the domination of the tensile energy component $E_s$ and the slight deviation comes from torsional strain components $E_\theta$ for most twisted AGNRs.

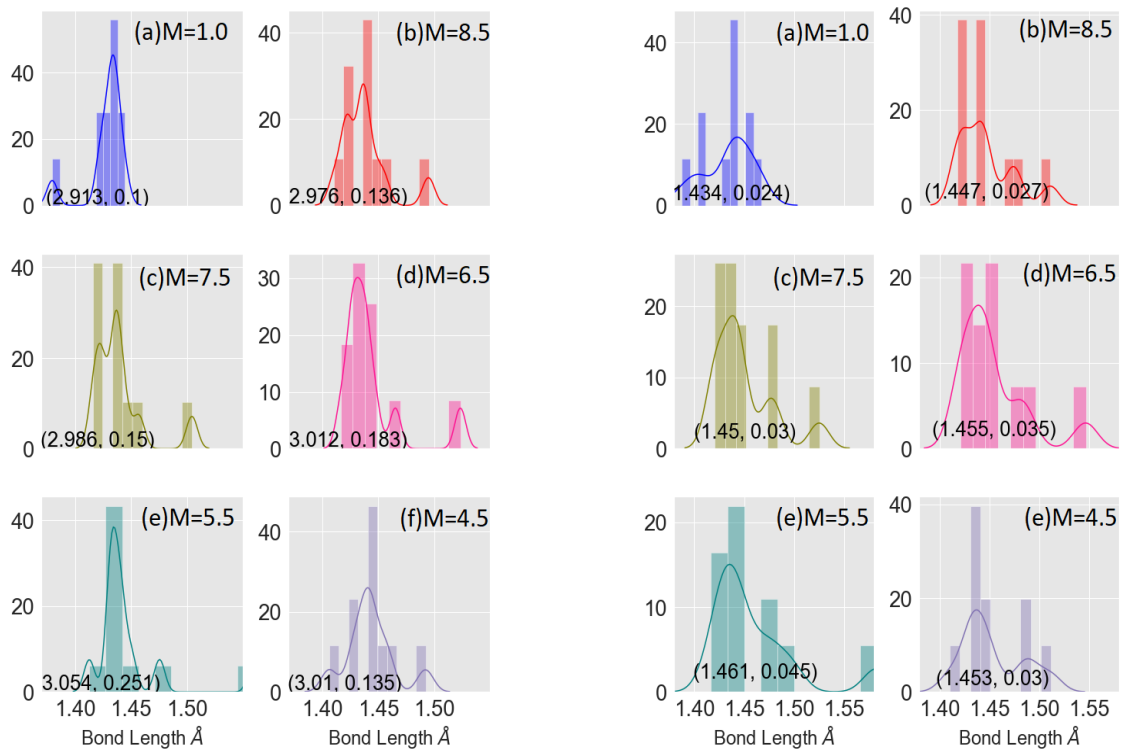

**Figure 5:** *Variation of normalized bond length as a function of torsional angle (a) $\theta = Zero$ (b) $\theta = 0.091 rad$ (c) $\theta = 0.104 rad$ (d) $\theta = 0.121 rad$ (e) $\theta = 0.146 rad$ and (f) $\theta = 0.182 rad$ for monolayer N=8 H-AGNRs. The first value inside the bracket is the average value of bonds and second value is the standard deviation in the bonds. Each bin represents the 0.0075Å width.*

**Figure 6:** *Variation of normalized bond length as a function of torsional angle (a) $\theta = Zero$ (b) $\theta = 0.084 rad$ (c) $\theta = 0.95 rad$ (d) $\theta = 0.110 rad$ (e) $\theta = 0.129 rad$ and (f) $\theta = 0.158 rad$ for monolayer N=8 F-AGNRs. The first value inside the bracket is the average value of bonds and second value is the standard deviation in the bonds. Each bin represents the 0.0075Å width.*

Furthermore, the energy necessary to deform the ribbons expressed as torsional energy per atom ($\Delta E/atom$), is lower for F-AGNRs than H-AGNRs for smaller values of $\theta$ and becomes higher in comparison to H-AGNRs as we increase $\theta$. For smaller $\theta$, the strain energy is negligible and lower value of ($\Delta E/atom$) shows the easiness of F-AGNRs to get twisted which need more torque for higher twist. The origin of this effect lies in the geometrical reconstruction due to the charge attained by passivated atoms at the edge as we will see later. As shown in the Figure 5(a-to-f) and Figure 6(a-to-f) the bond distribution of C-C bonds of various helical conformation that C-C bonds are not elongated homogenously for $N$=8 H & F passivated AGNRs as torsional strain increased in ascending order for (a) $M$=1.0 < (b) $M$=8.5 <(c) $M$=7.5 <(d) $M$=6.5 <(e) $M$=5.5 <(f) $M$=4.5, behavior which has also been reported in earlier studies[23,41,47,48].

Furthermore, the plasticity can be explained from the change of bond angles rather than bond lengths, therefore, the nonlinear variation can be explained by the bond angle alteration of benzo-rings. For comparison of bonds length response to torsional strain we can parameterize the effect of twisting on the bonds length of C-C of AGNRs as change in the mean value of C-C bonds length ($\bar{r}$) and standard deviation in the C-C bonds ($\Delta r$) due to the elongations of C-C bonds. The $\Delta r$ parameter for infinite graphene sheet must be equal to zero, however, ($\Delta r$) for pristine untwisted H & F passivated AGNR are not equal to zero because of slightly shorter bonds at edges[47] (but still AGNRs are bipartite and have very small value of $\Delta r$) as represented by the left most bin in Figure 5(a) and Figure 6(a). The bond distribution histogram shown in Figure 5(a-to-f) and Figure 6(a-to-f) for H & F passivated AGNRs shows us the marked upward shift of the shortest bonds that belongs to the edges, to left (longer region). This implies that, except most twisted ($M$=4.5) (Figure 5(f) and Figure 6(f)), for H & F passivated 8-AGNRs having $M$=1.0 to $M$=5.5, the peripheral C-C bonds experienced maximum stretch demonstrating an uneven distribution of strain evident from the successive increments of ($\bar{r}$) and ($\Delta r$) as shown in Figure 7 (See Table S1). However, in case of most twisted system $\theta$=0.162$rad$Å$^{-1}$ of 8-HAGNR and $\theta$=0.158$rad$Å$^{-1}$ of 8-FAGNR ($M$=4.5 for both cases) ($\bar{r}$) still observed to increase but ($\Delta r$) drops significantly implying the evenly distributed bond lengths to some extent. In this most rolled-up AGNRs because of high tensile strain and torsional strain the extent of overlapping between of the sp$^2$ orbitals get reduced and lower the stiffness of C-C bonds which is also reported in the nanotubes of smaller radii[47].

We can conclude that at extreme curvature (i.e. high value of $\theta$) AGNRs the C-C bonds are showing lesser stiffness are most likely unaffected from the passivation effect.

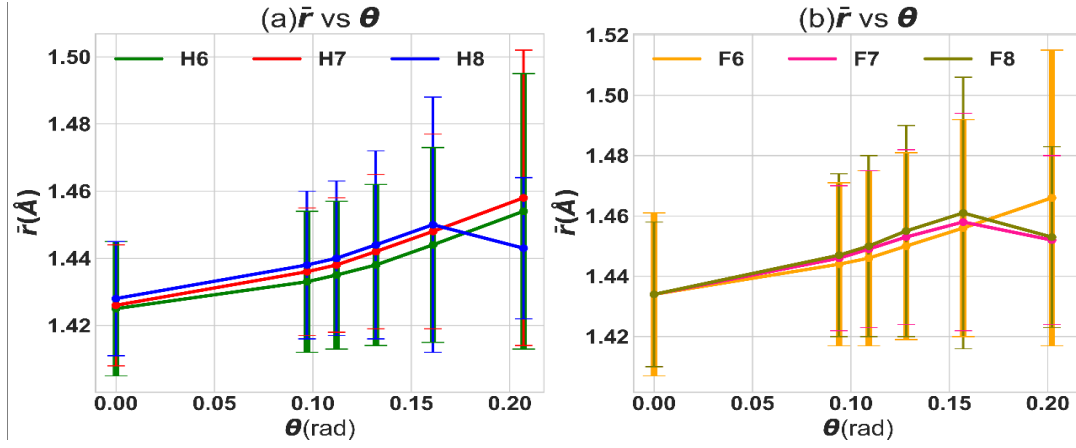

**Figure 7:** *Showing the effect of twisting on the average value of C-C bond length (solid circular marks) and the standard deviation in the distribution of bonds (vertical cap lines) of (a) H-AGNRs and (b) F-AGNRs.*

Furthermore, it is interesting to note that distribution curves of C-C bond length for H-AGRNs Figure 5(a-to-f) are narrower and its peak values go up to almost twice the value of the peak of much broader C-C bond distribution curve of F-AGNRs Figure 6(a-to-f), respectively. Broader distribution of peak and nonlinear $\sigma_z - \Sigma^2 \theta^2$ curves response for higher torsional strain implies the smoothness of bond angle alternation in case of twisted F-AGNRs.

The response could be understood qualitatively in terms of the tradeoff between the orbital overlap that corresponds to $E_\theta$ and the stiffness of C-C bonds. We can assume $\Delta E \simeq \mho \theta^4 \Sigma^4$, where $\mho$ is proportionality constant defined as rate of change of total energy w.r.t. the square of tensile strain. Furthermore, Hooke's Law may be written as $E_s \simeq \frac{C}{1-\nu^2} \theta^2 \Sigma^2$, where $C$ and $\nu$ is stiffness constant and Poisson's constant respectively then $\Delta E = E_s + E_\theta$, leads to equation:

$$E_\theta \simeq \mho \left( \theta^2 \Sigma^2 - C/2\mho(1-\nu^2) \right)^2 - C^2 / 4\mho(1-\nu^2)^2 \qquad (5)$$

Energetic analysis given in Equation (5) of twisted GNRs with open edges shows the torsional stress energy $E_\theta$ relation as a mechanical response of strain $\theta^2 \Sigma^2$ and may be applied even for wider AGNRs.

### 3.3. Electronic properties

We now consider electronic properties of these AGNRs. Note that the band gaps hierarchy for the known three families of planar AGNRs (defined by $q = mod\,(N, 3)$) depends on the width scaled as $E_g \approx \beta W^{-1}$, where $N$ is the numbers of dimers lines and $W$ is width[43,49]. Furthermore, in case of wider AGNRs the three families are barely distinguishable when subjected to torsional strain, but substantial changes occur only for the narrowest ribbons with $W <$ 1nm[50], therefore, has been taken in present study. Also, the sinusoidal response of band gap to the applied tensile strain[51] limits the control on strained engineered applications in nano-electronics[52,53]. In contrast, Figure 8(a) for AGNRs for $N$=6, 7 and 8 shows non-sinusoidal response of the band gap to torsional angle $\theta$. The trend for widening or narrowing the gap for q-dependent families defined as $q = mod(N, 3)$ which is associated with the number of dimers lines of AGNRs for $N$=6, 7 and 8. Although, the exception occurs for the widest $N$=8 HAGNR which as a quadratic fitting of parameters suggest narrowing of the gap for initial small values of $\theta$.

Except $N$=6 F-AGNRs, passivation does not change the trend actually but does change the zero order term significantly for different values of $N$ that shift the band lines downward. We can compare gentle bends in armchair ribbons that has been reported to cause significant widening or narrowing of energy gaps as $\Delta E_g \propto (-1)^q \theta^2$ for $q$=0 and 1 (here $\theta$ is the magnitude of small bending)[2]. In our study, the energy gaps for $q = 0$ and 1, H-AGNRs depends in $\theta$ space as $\Delta E_g \propto (-1)^q (\theta \delta + \theta^2 2^q \eta) + O(\theta)$ with $\delta = 2.7$ & $\eta = 6.8$, though, small $\theta^2$ makes quadratic contribution very small leading to almost linear dependence. For $q$=2, H-AGNRs the energy gap $E_g \propto (-1)^{q-1}(\theta\delta) + (-1)^q\theta^2\eta + O(\theta)$ with $\delta = 2.7$ again and $\eta = 44.2$ now $\eta$ is no longer insignificant for small values of $\theta$ and narrows the gap for smaller values of $\theta$ (dotted blue line in Figure 8(a)). Because of ambiguous dependence (i.e. non-sinusoidal but also non-monotonous) of $q$ in torsional angle space we transformed our current space to effective tensile strain space in search for better relation.

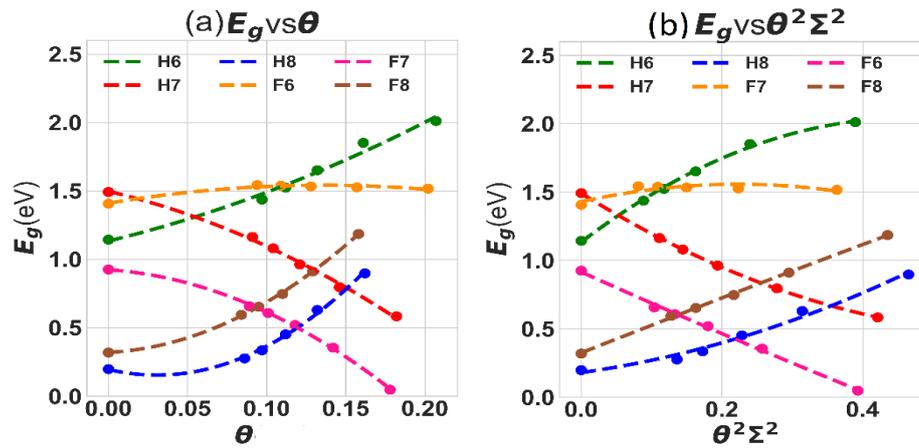

**Figure 8:** *(a) Non-linear response of band gap $E_g$ to torsional angle ($\theta$). N=8 HAGNR (blue dotted line) also showing non-monotonous response. (b) Linear as well as monotonous behavior of band gap $E_g$ in effective tensile strain ($\theta^2\Sigma^2$) space.*

So next, as a response of the effective torsional strain ($\theta^2\Sigma^2$) experienced by the AGNRs in Figure 8(b) shows the trend of band gap as $q = mod(N, 3)$ dependent which is governed by:

$$\Delta E_g \propto (-1)^{q-1}\delta(\theta^2\Sigma^2) + \beta O(\theta^2\Sigma^2) \quad (6)$$

where $O(\theta^2\Sigma^2)$ is the second order values which is insignificant for $N$=7 and 8 leads to the linear response of band gap i.e. the monotonous linear increase for $N$=7 ($q$=1) and monotonous linear decrease for $N$=8 ($q$=2) for both types of AGNRs (H-AGNRs and F-AGNRs) subject to $q$ dependence in tensile strain space. In case of $N$=6 ($q$=0) H-AGNRs, the second order term has a small effect though the band gap response remains monotonically increasing. Whereas, in case of *N*=6 F-AGNRs, it remains true for lower values of torsional strain ($\theta^2 \Sigma^2$) but deviates for higher values because its tendency for homogeneous distribution of strain amongst C-C bonds that reduced the inverse Poynting stress Figure 4(b). However, in case of twisted AGNRs we have not found any DFT study for comparison with our results except the study done by Zhang et al[41] using density functional tight binding method in which the band gap response for H-AGNRs to effective strain came out to be strictly linear in contrast to our study which shows it's not always the case. But for sake of relevancy we present a comparison of our study in Table 3 with other available experimental and theoretical studies for untwisted H-AGNRs as we will discuss it later.

Because the response of band gap follow an identical trend, the Equation (6) for $\Delta E_g$ described response for *N*=7 H-AGNR and *N*=8 H-AGNR remains the same for *N*=7 F-AGNR and *N*=8 F-AGNRs, respectively, except the shift in zero order term is required. Gap $E_g$ remains least responsive for $\theta$ in case of *N*=6 F-AGNR. Because of electronegative nature of F atom the charge shifted from narrow width AGNR strip to F atoms significantly depletes the electron density form whole ribbon, but does the distortion in shape of AGNRs change the retaining capacity of electron of AGNR? To answer this question we have applied the Hirshfeld method[54] to calculated the net charge accumulated on the H **&** F passivated atoms which is respectively $\Delta Q$= +0.02e and $\Delta Q$= -0.03e that does not vary on twisting and insignificantly vary ($\simeq |0.002|$) on varying dimer number *N*. Interestingly, in comparison to *N*=7 H-AGNRs, Figure 8(a) (red dashed line) the *N*=7 F-AGNRs, Figure 8(a) (pink dashed line) shows the passivation effect markedly, where the gap get narrowed down close to the 0.05eV for F-AGNRs forming a Dirac cone at $\pm K$ (Figure S4) as $\theta$ goes to its extreme, allowing ballistic transport of electrons of high kinetic energy at low bias.

### 3.4. Comparison of Present Study with Other Studies

It is evident from the Table 3 that theoretical and experimental band gap values of untwisted AGNRs for some cases differ significantly. Here we discuss the reasons for such differences and what the present study points to and the predictions it makes.

Table 3 *Measured and Calculated band gaps (in eV) for the untwisted pristine N=6, 7 and 8AGNRs*.

| 6-HGNARs | 7-HGNARs | 8-HGNARs | |
|---|---|---|---|
| **1.11** | **1.66** | **0.43** | ref[55] PBE |
| **1.12** | - | **0.20** | ref[42] PBE |
| **1.34***, **1.02**[#] | 1.27*, **1.57**[#] | 0.01*, **0.25**[#] | ref[43], TB*, LDA[#] |
| - | 2.3 | - | ref[56] TDFT-PBE0 |
| **1.12***, 2.68[#] | **1.62***, 3.81[#] | **0.30***, 1.15[#] | ref[44] LDA*, GW[#] |
| - | 1.6 | - | ref[45] Exp. |
| - | 2.5±0.1 | - | ref[57] Exp. |
| - | 2.3±0.1*, 3.7[#] | - | ref[58] Exp.*, GW[#] |
| - | 2.7 | - | ref[59] Exp. |
| - | 2.62*, **1.67**[#] | - | ref[60] Exp.*, PBE[#] |
| 1.14 | 1.49 | 0.20 | *Our Results PBE* |

Let us consider experiment versus experiment variation first. The band gap of 7-HAGNRs deposited on Au[111] substrate measured in ref [45] is 1.6eV which is significantly smaller than the later reported values 2.5eV, 2.6eV and 2.7eV in ref [57], ref[58] and ref[59] respectively. This variation in the band gap is likely to be there because of the difference in the reported apparent height of AGNRs which is 1.8Å in ref [45], 2.1 ± 0.1Å ref [57], 3.15Å in ref [58] and 1.85 ± 0.12 Å in ref [59]. Height of AGNR from surface is a crucial factor that has been studied theoretically in ref [56] which decides the electron-electron interaction at the interface. Furthermore, the length of synthesized AGNRs, details of which have not been given explicitly in any experimental study, can also be the reason for the variation as Zdetsis. at. el.

suggested theoretically[56]. Besides the variation in apparent height and AGNRs' length the slight charge transfer to substrate can also alter the band gap drastically[61].

In theoretical findings the band gap values calculated at GGA-PBE level of theory as in our study, are in excellent agreements with earlier study in ref [42,55]. However, as shown in Table 3 the band gap of 7-HAGNRs calculated using GW functional is ~2.5 times higher than the band gap calculated by using PBE functional. The band gap of 7-HAGNRs calculated using GW functional is overestimated by ~42%[44,57–59] and the band gap value calculated by using PBE functional is underestimated by ~40%[43,55,60] from the experimental band gap. As mentioned earlier, Zdetsis. at. el. demonstrated theoretically that besides the width, the length and the length-variation is an important factor and has been overlooked in previous literature[56].

In order to check substrate effect on AGNRs, further experimental as well as theoretical studies with different substrates need to be done. The possibility of the effect of the substrate is strongly suggested by theoretical study in which Romaner et. al. have observed the class of π-conjugated molecules (3,4,9,10-perylene-tetracarboxylic acid dianhydride (PTCDA)) adsorbed on the Ag(111), Au(111) or Cu(111) surfaces, shows not only the characteristic trends for work-function modification but also a net metal-to-molecule electron transfer[61]. Note that in our study, a strong sensitivity of band gap to the polarization due to the charge transfer to edges atoms has been observed.

Our calculations, despite underestimations of the band gap, still hold relevance for understanding the behavior and trend of band gap response to effective torsional strain and the mechanical properties of free standing long AGNRs or staircase AGNRs which have been synthesized with bottom-up approach. In this study we have also parameterized the effect of strain on the structure of helical shaped AGNRs that add up to the details of system and can be compared quantitatively with any future study.

## 4. Summary

In summary, we have performed DFT based calculations to demonstrate the symmetry and torsional strain effect on the severely twisted helical morphologies of H-AGNRs and F-AGNRs with fixed end configuration.

In the induced tensile strain-stress space, the linearity of tensile strain to axial tensile stress shows super elasticity in the case of H-AGNRs even at high torsional strain, while F-AGNRs shows plastic deformation at higher strain because of higher torque required to twist the ribbon after some threshold value of torsional strain is reached. System under the non-elastic region shows softening and enter in a plastic state, which may be easily destroyed by vacancy defects, long wavelength perturbations and high temperature effects[62]. This plastic deformation is understandable as a benzo-ring deformation due to the bond angle alternation and also as a tradeoff for orbital misalignment along the axial C-C bonds with its stiffness. Not much effect of symmetry is observed for mechanical properties on twisted ribbons.

Furthermore, the $q$ dependence of AGNRs on the electronic properties of helical conformations of narrow AGNRs for $N$=6, 7 and 8 can further divided in two groups for severely twisted and strained helical conformations as $q$=1 & 2 and $q$=1, in effective tensile strain $\theta^2 \Sigma^2$ space. Band gap is linearly and monotonously increasing for $q$=0 & 2 or decreasing for $q$=1 as a response against effective tensile strain. This trend remains unchanged for F-AGNRs, though because of zero order downward shift the band gap drops more to $E_g \simeq$ 0.05$eV$ at torsional strain $\theta$=0.178$rad$Å$^{-1}$ forming Dirac cone at $\pm$K. Also, net charge on the passivated atoms remains unaltered on twisting or increasing the number of dimers. Hence, we propose a *q* dependent monotonous and linear trend of electronic gap for twisted and strained helical conformations of bipartite lattice of narrow **N**=6, 7 & 8 H-AGNRs and F-AGNRs.

## Acknowledgements

High performance computing facility of Centre for Development of Advanced Computing (C-DAC), Pune and CVRAMAN, high performance computing cluster, at Himachal Pradesh University, Shimla has been used in obtaining the results presented in this paper. Author acknowledge the SIESTA TEAM for providing code under free licence. RT acknowledge special thanks to Professor Ahluwalia for proofreading and critique of this paper.

x.